# Embedding Radiomics into Vision Transformers for Multimodal Medical Image Classification

Zhenyu Yang[1,2,†,*], Haiming Zhu[1,2,3,†], Rihui Zhang[1,2,†], Haipeng Zhang[1,2,4], Jianliang Wang[2,4], Minbin Chen[2,5], Fang-Fang Yin[1,2], and Chunhao Wang[4]

1. Medical Physics Graduate Program, Duke Kunshan University, Kunshan, Jiangsu, China
2. Jiangsu Provincial University Key (Construction) Laboratory for Smart Diagnosis and Treatment of Lung Cancer, Kunshan, Jiangsu, China
3. Department of Radiotherapy and Oncology, The First People's Hospital of Kunshan, Kunshan, Jiangsu, China
4. Department of Radiation Oncology, Duke University, Durham, NC, United States
5. Department of Radiology, The First People's Hospital of Kunshan, Kunshan, Jiangsu, China

[†] Equally Contributed

*Corresponding author:
Zhenyu Yang, Ph.D.
Medical Physics Graduate Program
Duke Kunshan University
Kunshan, Jiangsu, 215316, China
E-mail: zhenyu.yang@dukekunshan.edu

**Short Running Title: Embedding Radiomics into ViT**

# Abstract

**Background**: Medical image analysis has witnessed substantial advancements through recent deep learning (DL) algorithms development. Vision Transformers (ViTs) have emerged as a powerful solution alternative by leveraging self-attention to model both local and global interactions. Despite their promise, ViTs are data-intensive and lack inductive biases, limiting their utility in medical imaging. Conversely, radiomics offers domain-specific, interpretable descriptors of image heterogeneity but lacks scalability and integration with deep learning. This study proposes a unified Radiomics-Embedded Vision Transformer (RE-ViT) framework that combines handcrafted radiomic features and data-driven visual embeddings within a ViT architecture.

**Purpose**: To develop and evaluate a hybrid RE-ViT framework that integrates radiomics and patch-wise ViT embeddings through early fusion, enhancing robustness and performance in multimodal medical image classification.

**Methods**: Following the classic ViT design, input image was first resampled into multiple image patches. For each image patch, handcrafted radiomic features, including intensity, texture, and spatial heterogeneity descriptors, were extracted. Simultaneously, standard patch embeddings were obtained via linear projection of pixel intensities. The two embeddings were averaged, normalized, and combined with positional encodings before being tokenized and processed by a ViT encoder. A learnable token aggregates patch-level information for final classification. The model was evaluated on three publicly available datasets, BUSI (lesion malignancy diagnosis through breast ultrasound), ChestXray2017 (lung pneumonitis diagnosis through chest X-ray), and Retinal OCT (retina disease diagnosis through retinal OCT), using 10-fold cross-validation. Performance metrics included accuracy, macro area under the ROC curve (AUC), sensitivity, and specificity. Ablation studies were implemented to assess the contribution of RE-ViT architectural components on these three clinical probelem. Comparative analyses were also conducted against CNN (VGG-16, ResNet) and hybrid (TransMed) models.

**Results:** The pretrained RE-ViT achieved state-of-the-art performance across all datasets. In BUSI, it achieved an accuracy of $0.848\pm0.027$, AUC of $0.950\pm0.011$, sensitivity of $0.796\pm0.042$, and specificity of $0.905\pm0.020$. In ChestXray2017, it yielded an AUC of $0.989\pm0.004$ and sensitivity of $0.953\pm0.010$, outperforming all other models. In Retinal OCT, RE-ViT achieved an AUC of $0.986\pm0.001$ and sensitivity of $0.914\pm0.023$. In the comparison studies, the proposed RE-ViT consistently matches or outperforms alternatives. Ablation revealed significant performance drops when removing either radiomics or projection-based embeddings. Attention map visualizations demonstrated modality-specific utilization of radiomics and learned features, with improved localization of clinically relevant regions.

**Conclusions:** The proposed radiomics-embedded vision transformer was successfully developed for multiple image classification tasks. Current results underscore the potential of our approach to advance other transformer-based medical image classification tasks.

# Introduction

Medical imaging plays a pivotal role in modern diagnostics and clinical decision-making, offering non-invasive visualization of anatomical structures and pathological processes across a wide range of diseases [1]. With the advancement of machine learning, particularly deep learning, numerous automated image analysis methods have been developed to assist in improving diagnostic accuracy and efficiency [2]. Among these, convolutional neural networks (CNNs) have been widely adopted due to their hierarchical architecture and ability to effectively capture local spatial features [1], [3], [4], [5]. CNN-based approaches have achieved substantial success across various medical image analysis tasks. Despite these achievements, CNNs are inherently constrained by their localized convolutional kernels, which can limit their capacity to capture long-range contextual information [6]. This can be important in applications where spatially distributed pathological signals are diagnostically relevant. In this context, Vision Transformers (ViTs) have emerged as an alternative deep learning architecture [2], [7], [8]. By dividing an image into patches, projecting each into a high-dimensional space, and modeling their interactions through a self-attention mechanism, ViTs provide a flexible means of learning both local and global dependencies. Their effectiveness has been demonstrated in natural image analysis tasks, such as classification, detection, and segmentation [7], [8]. However, the application of ViTs to medical imaging remains relatively limited. Their high data requirements and absence of inherent inductive biases—such as spatial locality and translation invariance, pose challenges in clinical domains where annotated datasets are often small. Moreover, ViTs trained from scratch may not effectively capture clinically meaningful patterns, and their purely data-driven nature can make them less interpretable and harder to align with expert knowledge [7].

In parallel, radiomics has gained attention as a complementary strategy [9], [10]. It involves the extraction of quantitative features (such as intensity statistics, shape descriptors, and texture patterns) from regions of interest in medical images. These features have shown associations with histopathological findings, genetic markers, and clinical outcomes across different modalities. For example, in breast ultrasound imaging, radiomic features have been extensively used to characterize tumor heterogeneity, which correlates with histological grade, malignancy risk, and treatment response [11], [12]. Studies have consistently demonstrated the efficacy of radiomics in differentiating between benign and malignant breast masses [11]. In chest radiography, radiomics has been utilized to quantify pulmonary abnormalities associated with pneumonia, chronic obstructive pulmonary disease, and tuberculosis, with particular success in evaluating disease severity and predicting therapeutic outcomes [13], [14]. The application of radiomics has also expanded into ophthalmic imaging [15]. In retinal Optical Coherence Tomography (OCT), radiomics has enabled the quantitative analysis of pathological changes in retinal layers, including fluid accumulation and structural distortions, across diseases such as diabetic macular edema, age-related macular degeneration, and central serous chorioretinopathy [16]. While interpretable, this modular design hinders end-to-end optimization and often requires manual feature engineering tailored to specific imaging modalities or anatomical sites. As a result, the pipeline is not well-suited for large-scale or heterogeneous datasets and exhibits limited adaptability to task-specific variations or domain shifts [17], [18].

Recognizing the complementary strengths of radiomics and Transformer-based models, recent studies have explored the integration of these approaches [19], [20], [21]. Most existing methods employ late-fusion or post-hoc integration strategies. A common practice is to compute Transformer-derived attention, such as risk scores or feature vectors, and combine them with radiomics features using traditional machine learning models, such as logistic regression or LASSO, for downstream classification or prognosis prediction [22], [23]. This method treats radiomics as an external descriptor, separate from the core representation learning process of the Transformer. Another strategy involves extracting attention maps from trained Transformers to identify salient regions, from which radiomics features are then computed and fed into a separate classifier [19], [21]. While this approach attempts to guide feature extraction using learned attention, the handcrafted features remain decoupled from the attention mechanism itself. These frameworks fail to achieve joint end-to-end representation learning between radiomics and Transformer. As a result, the potential synergy between radiomics and attention-based modeling has not been fully exploited.

In this study, we propose a Radiomics-Embedded Vision Transformer (RE-ViT) framework, a hybrid architecture that integrates handcrafted radiomics features and patch-wise visual embeddings within a unified Transformer model. The framework adopts an early-fusion strategy, where radiomic features are embedded at the same spatial granularity as ViT image patches. Specifically, RE-ViT constructs two parallel embedding streams: one based on standard patch-wise linear projections, as in the original ViT design, and the other derived from radiomics features extracted from the corresponding image patches. These two streams are fused and tokenized, then passed to a Transformer encoder, which models both intra- and inter-patch dependencies through self-attention. This design allows the Transformer to capture interactions between handcrafted and learned features from the beginning of training. To assess the performance of RE-ViT, we conducted comprehensive experiments on three clinical tasks: breast lesion malignancy classification (normal, benign, malignant) based on ultrasound, the lung pneumonia detection (normal, bacterial pneumonia) based on chest X-ray, and retinal disease classification (normal, choroidal neovascularization, diabetic macular edema, and drusen) based on OCT. Systematic comparisons were carried out against established CNN architectures (VGG-16, ResNet), hybrid models (TransMed), and multiple ablation variants of RE-ViT to examine the contributions of individual components.

# Material and Methods

**A. Datasets**

Three publicly available medical image datasets were employed in this study with distinct imaging modalities and classification tasks: (1) the Breast Ultrasound Images (BUSI) dataset [24], (2) the ChestX-ray2017 dataset [25], and (3) the Retinal OCT dataset [25]. Table I summaries the sample number for these datasets. Specifically,

- The BUSI dataset is one of a representative dataset in breast imaging research for lesion detection, which comprises a total of 830 breast ultrasound images collected from 600 female patients from a single institution, with ages ranging from 25 to 75 years. The dataset includes three diagnostic categories: normal, benign, and malignant, with class-wise distributions of 133, 487, and 210 images, respectively. Each image was acquired using a General Electric LOGIQ E9 ultrasound system and was subsequently resampled to a 500 × 500-pixel size. All diagnosis labels were confirmed by board-certified radiologists.
- The Chest X-ray 2017 dataset, also derived from a single institution, provides critical radiological patterns associated with inflammatory lung conditions. The dataset is a widely accepted benchmark for evaluating deep learning models in recognizing pneumonia subtypes based on chest imaging. It contains a total of 5826 chest X-ray images, categorized into three diagnostic groups: normal (1,583 images), bacterial pneumonia (2,780 images), and viral pneumonia (1,493 images).
- The multi-institutional Retinal OCT dataset comprises 108,312 high-resolution retinal optical coherence tomography images collected from patients with various retinal pathologies. The dataset includes four clinically significant categories: normal (51,140 images from 3,548 patients, mean age 60 years, 59.2% male), choroidal neovascularization (CNV) (37,206 images from 791 patients, mean age 83 years, 54.2% male), diabetic macular edema (DME) (11,349 images from 709 patients, mean age 57 years, 38.3% male), and drusen (8,617 images from 713 patients, mean age 82 years, 44.4% male).

These datasets are widely used in the literature for benchmarking medical image classification models [5], [7], [26], [27].

*Table I. Summary of datasets included in this work.*

| Datasets | Class 1 | Class 2 | Class 3 | Class 4 |
| --- | --- | --- | --- | --- |
| **BUSI** | Normal (133) | Benign (487) | Malignant (210) | / |
| **ChestXray2017** | No Pneumonitis (2780) | Bacterial Pneumonitis (4193) | Viral Pneumonitis (5183) | / |
| **Retinal OCT** | Normal (51140) | CNV (37260) | DME (11349) | Drusen (5183) |

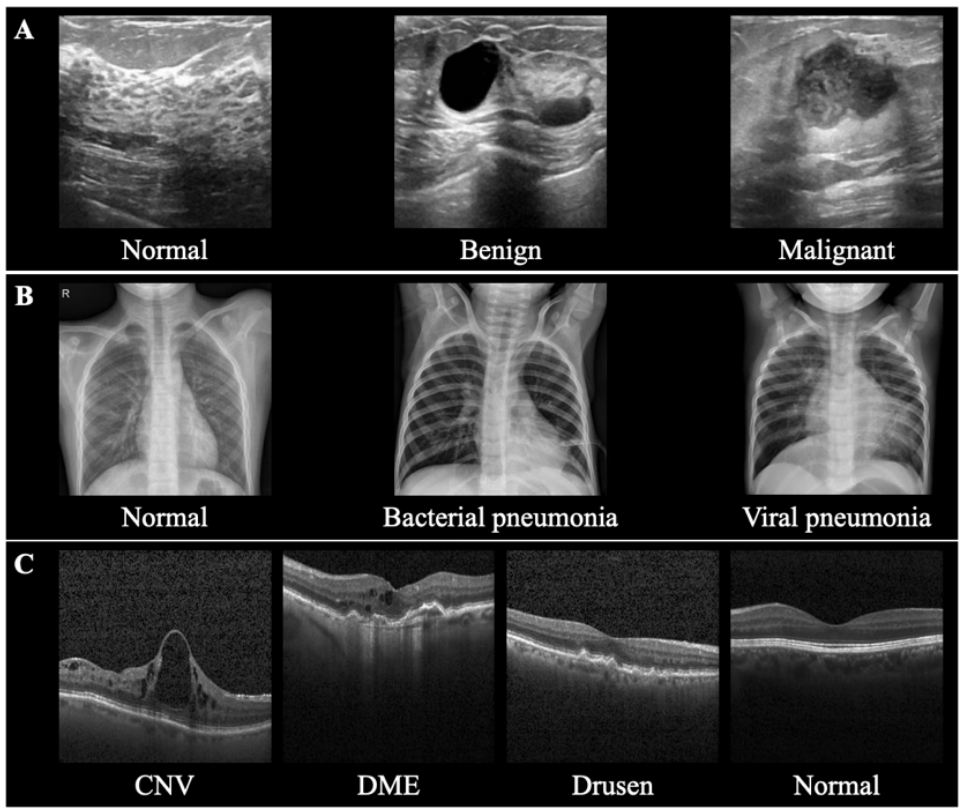

*Figure 1. The sample images of (A) Breast ultrasound image (BUSI) dataset, (B) Chest x-ray dataset 2017, (C) Retinal OCT dataset. The BUSI includes three categories: normal, benign, and malignant; the Chest x-ray dataset 2017 includes three categories: normal, bacterial pneumonia, and viral pneumonia; the Retinal OCT dataset includes four categories: CNV, DME, Drusen, and normal.*

## B. RE-ViT Model Design

The overall design of the Radiomics-Enhanced Vision Transformer (RE-ViT) model is shown in Figure 2. The architecture includes four key components: (1) radiomics-based embedding stream, (2) linear projection-based patch embedding stream, (3) positional embedding and token fusion, and (4) Transformer encoder for produce the final image classification. As shown in Figure 2(A), the input image is first partitioned into $N$ non-overlapping patches of size 16×16 pixels, following the standard ViT formulation [28]. Each patch is independently processed through both radiomics-based and projection-based embedding streams, after which the fused token sequence—including a prepended class token—is passed to the Transformer encoder to produce the final classification output [28].

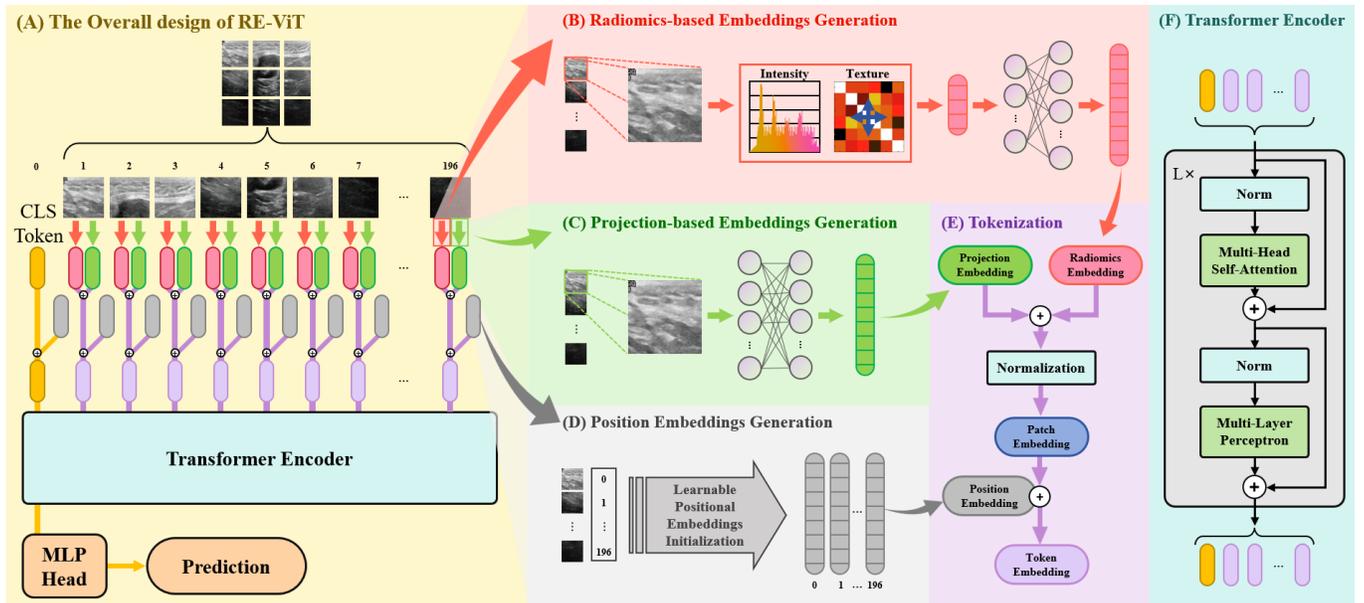

*Figure 2.* (A) The overall design of the RE-ViT model, which includes 4 steps: (B) radiomics embedding workflow that learns the image feature by using habitant radiomic analysis, (C) patch embedding workflow that learns the image feature by using linear projection-based methods, (D) position embedding workflow that providing spatial information, (E) combination of radiomics embedding, patch embedding, and position embedding results and tokenization, (F) transformer encoder for the final image classification.

*B.1. Radiomics-based Embedding*

The radiomics-based embedding stream introduces domain-specific, handcrafted features into the Transformer pipeline. As shown in Figure 2(B), a total of 91 radiomic features were extracted from each 16 × 16 patch. These features were selected to comprehensive characterize the locoregional image intensity and texture [29], [30], [31], which included 18 intensity-based features, 22 gray-level co-occurrence matrix (GLCOM)-based features, 16 gray-level run-length matrix (GLRLM)-based features, 16 gray-level size-zone matrix (GLSZM)-based features, 14 gray-level dependence matrix (GLDM)-based features, and 5 neighborhood gray-tone difference matrix (NGTDM)-based features.

According to standard radiomic processing pipelines, the intensity-based features were calculated directly from raw pixel intensities, while texture features (GLCOM-based, GLRLM-based, GLSZM-based, GLDM-based, NGTDM-based) were derived from 64-bin discretized images using a fixed-bin quantization scheme. To ensure consistency across the dataset, all extracted features are Z-score normalized. For each image patch, the obtained 91-dimensional radiomics vector was passed through a fully connected layer to map it into a 768-dimensional embedding space, aligning with the patch embedding dimension used in ViT. This projection enables the model to learn optimal linear combinations of handcrafted features, enhancing their semantic utility within the Transformer.

*B.2. Linear Projection-based Embedding*

Parallel to the radiomics-based embedding, the model employed a data-driven linear-projection embedding stream consistent with the original ViT design. As illustrated in Figure 2(C), each 16 × 16 image patch was first flattened into a 256-dimensional vector of raw pixel intensities, Similarly, the obtained feature vector was then projected into a 768-dimensional embedding via a learnable linear transformation. This embedding captures fine-grained, low-level spatial and intensity patterns, providing a complementary representation to the radiomics-derived features.

*B.3. Positional Embedding and Token Fusion*

Both radiomics and projection-based embeddings are independent of spatial location; hence, a learnable positional embedding is needed to retain spatial information among patches [28], [32]. As shown in Figure 2(D), a 768-dimensional positional vector $pos_i$ was generated for each patch index $i$. The final patch embedding for patch *i* was computed as:

$$x_i = \text{Norm}(\frac{rad_i + poj_i}{2}) + pos_i \qquad (1)$$

where $rad_i$ and $poj_i$ denote the radiomics-based embedding and projection-based embedding for the $i$-th patch, respectively. The embeddings are averaged and normalized using layer normalization to ensure uniform scaling. The fusion technique assumes both handcrafted embeddings and data-driven embeddings stream have equal initial importance, and the model adaptively reweight their contributions during training. The full token sequence $X = \{x_1, x_2, ..., x_N\}$, where $N$ is the number of patches, is constructed by concatenating all fused patch tokens. A learnable class embedding, a.k.a [CLS] token, is prepended to the sequence $X$ to aggregates information across all patches through attention interactions and, ultimately, serves as the final representation for classification [28].

*B.4. Transformer Encoder implementation*

The token sequence was then processed by a standard ViT encoder with $L$ stacked blocks. As shown in Figure 2(F), each block contains a multi-head self-attention (MHSA) module followed by a position-wise feedforward network (FFN). Within each MHSA module, the input tokens were linearly transformed into queries $Q$, keys $K$, and values $V$, and the self-attention was computed as:

$$\text{Attention}(Q, K, V) = \text{softmax}\left(\frac{QK^T}{\sqrt{d_k}}\right) \times V \qquad (2)$$

where $d_k$ is the dimensionality of the key vectors. This attention mechanism allows each token to attend to all others in the sequence, enabling the model to capture both local and global contextual relationships. Multiple attention heads operate in parallel to capture diverse semantic relationships. Their outputs were then concatenated and linearly transformed back to the embedding dimension. The FFN consists of two fully connected layers with a non-linear GELU activation in between. This module performs position-wise transformations to enrich the token representations learned by attention. Each block (MHSA and FFN) was wrapped with residual connections and layer normalization to facilitate training stability and gradient propagation.

After $L$ layers of transformation, the final representation of the [CLS] token is extracted and passed through a linear classification head to predict the image-level class label. This mechanism enables the model to integrate information across all patches and generate a compact, informative global feature representation.

## C. Model Implementation and Evaluation

The proposed RE-ViT model was implemented and evaluated under two training paradigms: (1) training from scratch, and (2) training with pretrained weights. Pretraining is widely regarded as essential for Vision Transformer models, particularly in medical imaging analysis, where annotated datasets are often limited, and the tasks require sophisticated spatial reasoning [33]. Standard ViT architectures are typically pretrained on large-scale natural image dataset to learn generalized visual representations. In our study, RE-ViT's projection-based embedding stream adheres strictly to the original ViT design, rendering it fully compatible with existing pretrained ViT weights. Therefore, in the second training paradigm, we initialized the projection embedding module using weights from a ViT model pretrained on the ImageNet-21k dataset [34], which contains over 14 million images spanning approximately 21,000 object categories. All other components of RE-ViT (including the radiomics embedding module, positional encoding vectors, and Transformer encoder layers) were initialized randomly and trained end-to-end.

Model implementation utilized a 10-fold Monte Carlo cross-validation protocol across three independent datasets. In each fold, the data were randomly split at the patient level into training and testing sets in an 8:2 ratio, and this procedure was repeated ten times with different random seeds to ensure robust performance estimation. Evaluation metrics included classification accuracy, one-vs-rest (OVR) macro area under the ROC curve (AUC), OVR macro sensitivity, and OVR macro specificity. Performance scores from all 10 folds were averaged to report overall model effectiveness and robustness.

To further investigate the contributions and explainability of the radiomics-based versus projection-based embedding streams, we conducted an analysis of raw attention maps derived from the initial Transformer encoder block. Raw attention visualization is well-established in attention-based explainability literature as a straightforward method for examining how attention scores reflect the model's initial representation of input features prior to deeper compositional transformations [35], [36]. As previously mentioned, our RE-ViT architecture employed a [CLS] token to aggregate global contextual information. During self-attention computation, the model generates attention scores that quantify how strongly the [CLS] token attends to each patch embedding token. Therefore, the raw attention scores from the first Transformer block can be extracted by isolating attention weights from the [CLS] token to all spatial patch tokens. These scores quantitatively represent the model's initial allocation of representational focus on each image patch. To separately evaluate the impact of each embedding stream, we executed the RE-ViT model twice for each input image: first employing only the radiomics-based embedding and subsequently using only the projection-based embedding, with all other model parameters and conditions held constant. The raw attention between all token pairs was then calculated:

$$\text{Attention}_{raw} = \text{softmax}(Q_{CLS} \times \text{K}_{tokens}{}^T) \tag{3}$$

where $\text{Attention}_{raw}$ denotes the initial attention scores between the CLS token and all other input tokens, computed as the matrix product of the class token's query vector $Q_{CLS}$ and the transposed key matrix $\text{K}_{tokens}$. The resulting raw attention scores for each scenario were individually reshaped into two-dimensional spatial grids reflecting the original image patch arrangement. These attention maps were subsequently upsampled to match the original input image resolution using bilinear interpolation and superimposed onto the images as heatmaps for better visualization [35].

The proposed RE-ViT model was implemented using PyTorch v2.0. All employed radiomic features were extracted using the public PyRadiomics library. The PyRadiomics toolbox has been intensively studied against the Image Biomarker Standardization Initiative (IBSI) guidelines. The model training was conducted for a maximum of 300 epochs with a learning rate initialized at 0.001. The optimization loss was set to the categorical cross-entropy loss. To prevent overfitting and promote generalization, an early stopping criterion was employed: training was halted if no improvement was observed in validation performance for 50 consecutive epochs. All experiments were executed on a workstation equipped with an AMD Ryzen 9 5950X 16-core CPU (3.4 GHz), 96 GB of system RAM, and an NVIDIA RTX 2080 GPU (11 GB VRAM). The core code of the RE-ViT model implementation will be available at [depository will be released later].

### D. Ablation study

To assess the individual contributions of key architectural components within the RE-ViT framework, a series of ablation studies were conducted. These experiments involved systematically removing or modifying specific modules of the model to isolate their impact on performance across three medical image classification tasks. Specifically, three variants of RE-ViT were evaluated as follows:

1) In the first variant, the radiomics-based embedding stream was entirely removed. The model retained only the projection-based patch embedding and positional encoding, thereby reverting to a standard ViT architecture. The goal of this variant was to quantify the impact of eliminating handcrafted radiomics features (i.e., locoregional intensity, texture, and heterogeneity) on the model performance.
2) In the second variant, the linear projection-based embedding stream was removed. The model operated solely on radiomics-based embeddings and positional encodings. This setup assessed whether domain-informed radiomic features alone, without raw pixel intensity information from the image patches, could sufficiently guide the Transformer to learn discriminative representations for classification.
3) In the third variant, the Transformer encoder was replaced with a deep CNN comprising 36 repeated convolutional layers with pooling operations. The network depth and parameter count were selected to approximate the computational capacity of the original Transformer module. All other components, including the radiomics embedding, projection-based embedding, and positional encoding, were retained. This comparison was designed to isolate the advantages conferred by the Transformer's self-attention mechanism, particularly its ability to model global context and long-range dependencies.

All ablation variants were trained and evaluated under the same experimental protocol as the full RE-ViT model. This included identical data preprocessing pipelines, 10-fold Monte Carlo cross-validation, evaluation metrics (accuracy, one-vs-rest macro-AUC, macro sensitivity, and macro specificity), and training hyperparameters (epoch limit, learning rate, early stopping strategy). The obtained performance in three classification tasks was then compared to the RE-ViT model using *Student's t-test* with a significant level of 0.05.

**E. Comparison study**

The performance of the developed RE-ViT model was additionally compared to three other common deep learning models in medical image analysis:

1) VGG-16: VGG16[37] is a deep convolutional neural network that consists of five sequential convolutional blocks, each followed by a max-pooling layer to reduce spatial dimensions. The convolutional blocks progressively increase the number of feature channels, starting from 64 and doubling up to 512, allowing the network to learn increasingly abstract and complex features. After the convolutional layers, the network includes three fully connected layers, culminating in a softmax output layer for multi-class classification tasks.

2) ResNet-50: ResNet50[37] is a deep convolutional neural network composed of an initial convolutional layer and max-pooling, followed by four stages of residual blocks with a total of 50 layers. Each residual block contains a bottleneck architecture with three convolutional layers and a shortcut connection that enables identity mapping or dimensional adjustment. After feature extraction, the network applies global average pooling and a fully connected layer to produce the final classification output.

3) TransMed: TransMed[38] is a hybrid deep learning architecture designed for medical image classification, combining CNNs with Transformer encoders. The network first employs CNN layers to extract local spatial features and generate a compact feature map, which is then flattened and embedded as a sequence of tokens. These tokens are passed through a Transformer module to capture global contextual relationships, followed by a classification head that integrates both local and global information for the final prediction. TransMed represents a class of architectures that integrate CNNs and Transformers, but it does not incorporate structured domain priors such as radiomics into the core attention mechanism.

All models were trained and evaluated under identical experimental conditions, including data preprocessing, 10-fold Monte Carlo cross-validation, training/testing splits, and performance metrics (accuracy, macro-AUC, macro sensitivity, and macro specificity). The obtained performance in three classification tasks was then compared to the RE-ViT model using *Student's t-test* with a significant level of 0.05.

# Results

The proposed RE-ViT model demonstrated superior classification performance across all three medical imaging datasets, particularly with pretrained weights based on ImageNet-21k. To evaluate the relative contributions of its constituent components, we conducted a comprehensive ablation study, summarized in Table II. In parallel, we benchmarked RE-ViT against three established deep learning architectures (i.e., ResNet, VGG-16, and TransMed) under both pretrained and non-pretrained settings, with results reported in Table III.

In the BUSI dataset, the pretrained RE-ViT model achieved the highest performance across all metrics, with an accuracy of $0.848 \pm 0.027$, AUC of $0.950 \pm 0.011$, sensitivity of $0.796 \pm 0.042$, and specificity of $0.905 \pm 0.020$. The non-pretrained RE-ViT exhibited a notable performance drop, particularly in AUC ($0.804 \pm 0.025$) and sensitivity ($0.617 \pm 0.045$), which demonstrated the critical role of transfer learning in data-limited clinical scenarios. When the radiomics embedding stream was removed (variant 1), performance decreased to an AUC=$0.941 \pm 0.018$ in the pretrained setting and AUC=$0.733 \pm 0.038$ without pretraining, suggesting a substantial contribution of handcrafted radiomics features. Excluding the projection-based embedding stream (variant 2) yielded an AUC=$0.762 \pm 0.097$. The variant 1 and variant 2 together confirming the complementary role of visual pattern extraction. Replacing the Transformer encoder with a CNN (variant 3) further reduced the AUC to $0.631 \pm 0.047$. When compared with alternative models, RE-ViT outperformed ResNet (AUC=$0.848 \pm 0.027$), VGG-16 (AUC: $0.813 \pm 0.036$), and TransMed (AUC: $0.741 \pm 0.050$).

In the ChestXray2017 dataset, the pretrained RE-ViT achieved an accuracy of $0.950 \pm 0.012$ and the highest AUC of $0.989 \pm 0.004$ among all ablation studies. Interestingly, the non-pretrained RE-ViT model also performed competitively (accuracy=$0.954 \pm 0.005$ and AUC=$0.982 \pm 0.002$). The results suggests that a relatively large dataset size could potentially reduce performance dependence on external pretraining. Exclusion of the radiomics embedding stream (variant 1) resulted in a modest AUC reduction to $0.979 \pm 0.004$ (pretrained) and $0.974 \pm 0.004$ (non-pretrained), while exclusion of the linear projection-based embedding stream (variant 2) and Transformer module (variant 3) yielded AUCs of $0.975 \pm 0.004$ and $0.889 \pm 0.034$, respectively. Comparatively, VGG-16 achieved the highest non-Transformer results (accuracy=$0.963 \pm 0.006$ and AUC=$0.993 \pm 0.002$), marginally surpassing RE-ViT without statistically significance.

In the Retinal OCT dataset, the pretrained RE-ViT again outperformed all evaluated models, achieving an accuracy of $0.938 \pm 0.001$ and an AUC of $0.986 \pm 0.001$. The non-pretrained RE-ViT yielded substantially lower values (accuracy: $0.872 \pm 0.006$ and AUC: $0.949 \pm 0.001$). Exclusion the radiomics embedding stream (variant 1) led to only a marginal AUC drop (AUC=0.984±0.001pretrained, AUC=0.933±0.005 non-pretrained), while exclusion the linear projection-based embedding stream (variant 2) and Transformer (variant 3) decreased the AUCs to 0.921±0.016 and 0.914±0.012, respectively. Among comparative models, VGG-16 (AUC=0.984±0.002), ResNet (AUC=0.962±0.031), and TransMed (AUC=0.979±0.001) lower than the pretrained RE-ViT in AUC and sensitivity.

Figure 3 visualize the attention map for radiomic-based embedding stream and projection-based embedding stream across the three tasks. Each panel displays the original input image alongside two heatmaps corresponding to the attention distributions derived from two embedding streams, respectively. These attention maps were first normalized to [0, 1] and were interpolated to match the original input image resolution and superimposed onto the images as heatmaps for better visualization. Distinct attention patterns can be observed across the three datasets. In the BUSI and ChestXray2017 datasets, the radiomics-based embeddings produced more localized and concentrated attention within diagnostically relevant regions, while in the Retinal OCT dataset, the projection-based embeddings yielded more anatomically aligned and coherent attention distributions. These qualitative results suggest that the model selectively leverages different embedding modalities based on image characteristics, with varying attention behaviors depending on the dataset and imaging modality.

***Table II***. *Ten-fold cross-validation classification results (mean ± standard deviation) for ablation studies.*

| | | RE-ViT (Pretrained) | RE-ViT (Non-Pretrained) | Variant 1 (Pretrained) | Variant 1 (Non-Pretrained) | Variant 2 | Variant 3 |
|---|---|---|---|---|---|---|---|
| BUSI | Acc. | **0.848±0.027** | 0.682±0.039* | 0.832±0.023* | 0.612±0.040* | 0.611±0.070* | 0.585±0.028* |
| | AUC | **0.950±0.011** | 0.804±0.025* | 0.941±0.018* | 0.733±0.038* | 0.762±0.097* | 0.631±0.047* |
| | Sens. | **0.796±0.042** | 0.617±0.045* | 0.771±0.046* | 0.561±0.033* | 0.562±0.131* | 0.391±0.038* |
| | Spec. | **0.905±0.020** | 0.805±0.021* | 0.892±0.019* | 0.780±0.015* | 0.765±0.059* | 0.701±0.024* |
| ChestXray2017 | Acc. | 0.950±0.012 | **0.954±0.005** | 0.889±0.020* | 0.940±0.008* | 0.935±0.011* | 0.861±0.031* |
| | AUC | **0.989±0.004** | 0.982±0.002 | 0.979±0.004* | 0.974±0.004* | 0.975±0.004* | 0.889±0.034* |
| | Sens. | **0.953±0.010** | 0.951±0.005 | 0.903±0.016* | 0.938±0.008* | 0.930±0.011* | 0.843±0.041* |
| | Spec. | 0.975±0.005 | **0.975±0.003** | 0.948±0.009* | 0.968±0.004* | 0.964±0.006* | 0.922±0.020* |
| Retinal OCT | Acc. | 0.938±0.001 | 0.872±0.006* | **0.939±0.003** | 0.848±0.004* | 0.827±0.013* | 0.857±0.006* |
| | AUC | **0.986±0.001** | 0.949±0.001* | 0.984±0.001* | 0.933±0.005* | 0.921±0.016* | 0.914±0.012* |
| | Sens. | **0.914±0.023** | 0.798±0.070* | 0.903±0.003* | 0.708±0.024* | 0.880±0.011* | 0.880±0.010* |
| | Spec. | 0.969±0.024 | 0.922±0.069* | **0.980±0.000*** | 0.955±0.001* | 0.795±0.017* | 0.779±0.018* |

*"\*" marks the statistically significant difference compared to the pre-trained version of RE-ViT model.*

***Table III***. *Ten-fold cross-validation classification results (mean ± standard deviation) for comparison studies.*

| | | RE-ViT Pretrained | RE-ViT Non-Pretrained | ResNet | VGG-16 | TransMed |
|---|---|---|---|---|---|---|
| BUSI | Acc. | **0.848±0.027** | 0.682±0.039* | 0.729±0.036* | 0.813±0.036* | 0.741±0.050* |
| | AUC | **0.950±0.011** | 0.804±0.025* | 0.848±0.034* | 0.925±0.017* | 0.880±0.019* |
| | Sens. | 0.796±0.042 | 0.617±0.045* | 0.721±0.035* | **0.813±0.035*** | 0.689±0.048* |
| | Spec. | **0.905±0.020** | 0.805±0.021* | 0.856±0.019* | 0.899±0.017 | 0.847±0.022* |
| ChestXray2017 | Acc. | 0.950±0.012 | 0.954±0.005 | 0.931±0.014* | **0.963±0.006*** | 0.925±0.017* |
| | AUC | 0.989±0.004 | 0.982±0.002* | 0.964±0.016* | **0.993±0.002*** | 0.971±0.005* |
| | Sens. | 0.953±0.010 | 0.951±0.005 | 0.926±0.018* | **0.961±0.007*** | 0.927±0.011* |
| | Spec. | 0.975±0.005 | 0.975±0.003 | 0.964±0.008* | **0.980±0.003** | 0.962±0.007* |
| Retinal OCT | Acc. | 0.938±0.001 | 0.872±0.006* | 0.906±0.046* | **0.941±0.007** | 0.926±0.001* |
| | AUC | **0.986±0.001** | 0.949±0.001* | 0.962±0.031* | 0.984±0.002 | 0.979±0.001* |
| | Sens. | **0.914±0.023** | 0.798±0.070* | 0.871±0.026* | 0.904±0.013* | 0.877±0.005* |
| | Spec. | 0.969±0.024 | 0.922±0.069* | 0.967±0.018 | **0.979±0.003*** | 0.975±0.001* |

*"\*" marks the statistically significant difference compared to the pre-trained version of RE-ViT model.*

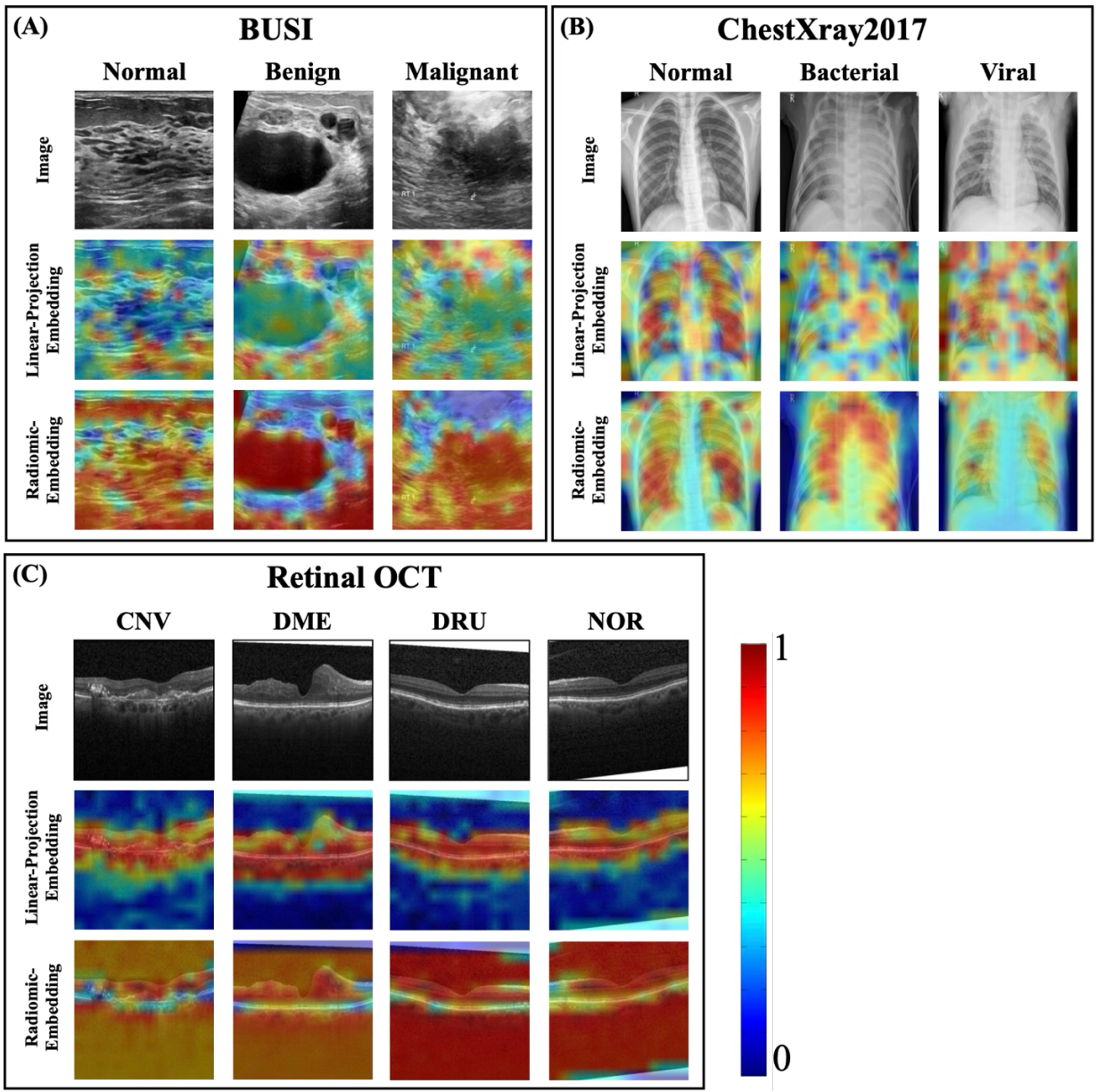

*Figure 3.* The attention map obtained for (A) BUSI classification, (B) ChestXray2017 classification, (C) Retinal OCT classification. Within each figure, the first row of images is the origin images, the second row is the attention map of linear projection-based embedding stream, and the third row is the attention map of linear projection-based embedding stream. These attention maps were subsequently interpolated to match the original input image resolution using bilinear interpolation and superimposed onto the images as heatmaps for better visualization.

# Discussion

The study introduced a novel radiomics-embedded ViT, a unified framework that strategically integrates handcrafted radiomic features and data-driven visual embeddings within a Transformer-based architecture. The classic Vision Transformers typically transformed image representation by segmenting input images into fixed-size patches, linearly projecting these patches into high-dimensional embeddings, and processing them through MHSA to model long-range spatial dependencies. Despite their success in various natural image analysis tasks, traditional ViTs face significant challenges in capturing medical images' subtle, localized pathological details due to their reliance on large, well-annotated datasets. Radiomics provides a complementary and clinically interpretable approach by extracting handcrafted descriptors related to image intensity, texture, and morphological patterns. These descriptors have consistently demonstrated clinical relevance and interpretability across diverse medical imaging modalities, effectively linking imaging phenotypes with underlying biological and pathological characteristics. RE-ViT strategically combining the complementary strengths of radiomics and ViTs, where the radiomic features were first extracted at the individual patch level. Such radiomic analysis is conceptually aligned with habitat radiomics, which emphasizes the extraction of localized intensity heterogeneity and textural patterns from distinct subregions. In parallel, projection-based embeddings directly encode raw pixel-level spatial and textural information. Both embedding streams are subsequently combined, tokenized and processed through a shared Transformer encoder, thereby simultaneously leveraging domain knowledge and data-driven visual pattern representation learning.

Extensive experiments on the three clinical problems (i.e., lesion malignancy diagnosis through breast ultrasound, lung pneumonitis diagnosis through chest X-ray, and retina disease diagnosis through retinal OCT) demonstrate that the proposed RE-ViT consistently matches or outperforms state-of-the-art CNN-based models (VGG-16, ResNet) and Transformer-based models (TransMed). Although conventional CNN models such as VGG-16 and ResNet performed competitively on certain tasks, particularly the ChestXray2017 dataset, their fixed architectures limit the integration of structured, domain-specific knowledge. Similarly, TransMed, which combines convolutional and Transformer components, demonstrated lower overall performance compared to RE-ViT. A prominent advantage of the RE-ViT architecture is its inherent compatibility with standard ViT frameworks, enabling straightforward integration of pretrained weights. The availability of pretrained weights (e.g., from ImageNet-21k) yielded substantial performance gains, particularly in the BUSI dataset (AUC=0.950±0.011 pretrained versus

AUC=0.804±0.025). The small dataset size and breast lesion variability posed significant challenges to training from scratch, which suggests the value of transfer learning in low-resource clinical domains. In contrast, the ChestXray2017 dataset contains a larger number of samples with more consistent anatomical patterns, which exhibited a minor difference between pretrained and non-pretrained configurations (AUC=0.989±0.004 pretrained versus AUC=0.982±0.002). The ablation experiments provided further insights into the contribution of individual architectural components. Replacement of the Transformer encoder with a deep CNN (variant 3) consistently degraded performance across all tasks (AUC=0.631±0.047, 0.889±0.034, 0.914±0.012 for BUSI, ChestXray2017, Retinal OCT dataset, respectively). The self-attention mechanism in Transformer encoder is critical in capturing long-range spatial relationships and complex interactions across the patches. Additionally, the removing either the radiomics (variant 1) or the projection-based embedding module (variant 2) individually resulted in noticeable performance degradation in all three tasks. Interestingly, for datasets with pronounced textural heterogeneity (BUSI and ChestXray2017), the radiomics-only variant achieved superior performance compared to the projection-only variant trained from scratch (AUC=0.762±0.097 versus AUC=0.733±0.038 for BUSI, and AUC=0.975±0.004 versus AUC=0.974±0.004 for ChestXray2017, respectively). This observation supports the efficacy of handcrafted radiomic features in the absence of extensive pretraining, particularly when discriminative pathological features are subtle and localized (as in BUSI).

The complementary advantages of the embedding streams were further illustrated through attention visualization maps derived from the early self-attention layers. Importantly, these raw attention scores are extracted from early layers (before any deeper compositional reasoning is performed) in RE-ViT, offering a more faithful representation of the model's initial inductive biases. The attention visualization in Figure 3 independently evaluates the attention behavior of the radiomics-based and projection-based embeddings. In BUSI, radiomics-based attention maps precisely localized tumor boundaries and internal heterogeneity, whereas linear projection-based maps were more diffuse and less discriminative. Similar trends were also observed in ChestXray2017, where radiomics-guided attention concentrated on regions of pulmonary opacity, showing better alignment with clinically relevant features. In contrast, in the Retinal OCT dataset, linear projection-based embeddings yielded more anatomically consistent attention maps that closely followed retinal layer boundaries and pathological deformations associated with CNV, DME, and DRU. Radiomics attention in OCT was less spatially coherent, indicating a modality-specific limitation of handcrafted features in capturing layered anatomical structure. These observations confirm the model's flexibility in selectively leveraging the most informative embedding modules depending on the imaging modality, thereby enhancing performance and robustness across anatomical contexts.

Despite these promising results, several limitations remain in the proposed RE-ViT model need to be addressed in future works. Although the employed datasets are diverse in modality, these data were restricted to public benchmarks and may not fully reflect the heterogeneity and complexity of real-world clinical environments. Further validation on large-scale, multi-institutional datasets is essential to assess generalizability and robustness in broader clinical deployment. Additionally, the radiomic features employed in this study were fixed and not tailored specifically to each imaging task, potentially limiting efficiency and introducing redundancy. Incorporating adaptive, task-specific radiomic feature selection or employing learnable radiomic representations could potentially further enhance model performance and efficiency. While attention map visualization provided valuable qualitative insights into model explainability, it does not offer quantitative explanations required for formal clinical validation. Integration with more rigorous interpretability frameworks, such as SHAP values [39], concept activation vectors [40], etc., could facilitate deeper quantitative assessment of model reasoning for clinical users.

# Conclusion

This study introduces RE-ViT, a novel radiomics-embedded Vision Transformer framework designed to integrate handcrafted semantic features with data-driven visual embeddings for medical image classification. The extensive evaluations and comparative analyses underscore the complementary roles of radiomics, projection-based embeddings, and Transformer encoders, each contributing to the model's robust and generalizable performance. By bridging the gap between domain-specific prior knowledge and modern deep learning architectures, RE-ViT not only advances the state-of-the-art in medical image analysis but also offers a scalable and interpretable solution to various medical image diagnosis tasks.